# Electric field induced irreversible transformation to the R3c phase and its influence on the giant piezo-strain behaviour of lead-free system- $Na_{0.5}Bi_{0.5}TiO_3$-$BaTiO_3$-$K_{0.5}Na_{0.5}NbO_3$

Dipak Kumar Khatua,[1] Anatoliy Senyshyn,[2] and Rajeev Ranjan[1*]

[1]Department of Materials Engineering, Indian Institute of Science, Bangalore-560012, India

[2]Forschungsneutronenquelle Heinz Maier-Leibnitz (FRM II). Technische Universität München, Lichtenbergestrasse 1, D-85747 Garching b. München, Germany

## Abstract

A systematic study of the structural, dielectric and piezoelectric behaviour of the lead-free system (0.94-x) $Na_{0.5}Bi_{05}TiO_3$-$0.06BaTiO_3$-$xK_{0.5}Na_{0.5}NbO_3$ exhibiting giant piezo-strain was carried out as a function of composition ($0.0 \leq x \leq 0.025$), electric field and temperature. While x-ray diffraction revealed cubic structure for all the compositions, neutron diffraction patterns shows weak superlattice peaks characteristic of a $\sqrt{2}$ x $\sqrt{2}$ x 16 type modulation. The giant piezostrain is associated with field induced transformation of this modulated structure to rhombohedral (R3c) phase. It is shown that the maximum achievable strain decreases substantially in subsequent cycles due to partial irreversibility of the field induced rhombohedral phase after the first cycle.

*rajeev@materials.iisc.ernet.in

The superior properties of Pb-based piezoceramics have dominated the sensor and actuator industry for the past many decades. However, environmental concerns associated with the toxicity of lead have led to increasing research activities on Pb-free piezoelectrics in the past one decade. $Na_{0.5}Bi_{0.5}TiO_3$ (NBT) based lead free piezoceramics have received considerable attention in this regard.[1-4] A giant strain of 0.45 % at electric field of 80 kV/cm was discovered by Zhang et al. in a ternary system $0.92Na_{0.5}Bi_{0.5}TiO_3$-$0.06BaTiO_3$-$0.02K_{0.5}Na_{0.5}NbO_3$ (0.02KNN).[3] Interest has therefore grown up on such ternary systems in the recent few years.[4-12] Jo et al.[4] suggested that the origin behind the giant strain response is existence of nonpolar phase that brings back the system to its original position once the applied electric field is removed. Zhang et al.[12] suggested that such a large strain is due to a field-induced antiferroelectric-ferroelectric phase transition. Using in situ electric field dependent TEM studies on 0.03KNN (following earlier notation for ref.3) Kling et al.[7] have shown that lamellar domain appears in the specimen as electric field is switched on and domain disappears as field switched off. They proposed origin of giant strain is due to field induced reversible rhombohedral phase transformation. X-Ray, Neutron powder diffraction and TEM study on 0.02KNN and 0.01KNN suggest an existence of almost non polar tetragonal phase with a space group P4bm and a rhombohedral phase R3c.[11] Hinterstein et al.[6] suggested that origin of large strain is due to field induced almost non polar tetragonal phase to rhombohedral phase transformation which is reversible in nature. Evidently, in spite of the several studies the understanding of this system's structure-property correlation is still incomplete. In this paper, we have carried out a systematic study of the structure-property correlation of this system as a function of composition, electric field and temperature. The study unravelled that the commonly reprted cubic structure of this system rather exhibits a long modulation period, and that giant strain diminishes after the first cycle due to partial quenching of the field induced rhombohedral state. The critical frequency below which the giant strain becomes manifested was 0.5 Hz.

Different compositions of $(0.94-x)Na_{0.5}Bi_{0.5}TiO_3$-$0.06BaTiO_3$-$xK_{0.5}Na_{0.5}NbO_3$ ($0 \leq x \leq 0.025$) were prepared following conventional solid state route by mixing stoichiometric ratios of dried $Na_2CO_3$ (99.9%), $Bi_2O_3$ (99%), $TiO_2$ (99.8%), $K_2CO_3$ (99.5), $BaCO_3$ (99%) and $Nb_2O_5$ (99.95% ) in an acetone medium using zirconia vials and balls in a planetary ball mill for 12 hrs. Calcination was carried out at 900 $^o$C for 3 hrs. After thorough ball milling, the calcined powder was recalcined at 900 $^o$C for 2 hrs. Green pellets were made first by applying uniaxial pressure of 100 MPa and then subjecting them to cold isostatic pressing at 300 MPa. Sintering was carried out at 1150 $^{\circ}$C for 3 hrs. The sintered pellets were found to be ~ 97.5 % dense. X-ray powder diffraction (XRD) was carried out with Cu-K$\alpha_1$ radiation using Rigaku Smartlab X-Ray diffractometer. Neutron powder diffraction (NPD) patterns were collected at FRM-II, Germany (wavelength 1.5483 Å). Electric poling was carried out by applying a dc field of 45 kV/cm for 30 minutes. Dielectric measurements were carried out on a Novocontrol impedance analyzer (Alpha-A). A precision Premier II tester (Radiant Technology, Inc.) was used for electric field-polarization (E-P) measurement. Piezoelectric coefficient $d_{33}$ was measured using a Berlincourt meter piezotest (model PM300). Electric field dependent strain measurement was performed using MTI-2100 FOTONIC SENSOR (mti instrument). Structural analysis was carried out by Rietveld method using the FULLPROF package.[13]

In conformity with the previous studies, X-ray powder diffraction patterns (XRPD) of all the compositions revealed cubic structure with the lattice parameter increasing with increasing KNN concentration. In contrast, NPD pattern of all the compositions show weak superlattice reflections. If these superlattice reflections are ignored the main Bragg peaks in the NPD suggests a cubic lattice in agreement with the XRD patterns. Interestingly, superlattice reflections have also been observed in electron diffraction studies. Kling et al.[7] have reported in their TEM study of 0.03KNN that unpoled single grain contain P4bm+R3c phase with high intensity of tetragonal (P4bm) superlattice reflections. In a subsequent report the authors suggested cubic + tetragonal + rhombohedral phase coexistence in unpoled state of 0.02KNN.[10] If the superlattice reflections in the neutron diffraction patterns originate from the same structural distortions, i.e. P4bm and R3c as reported by Kling et al., then the P4bm + R3c phase coexistence models is expected to fit the superlattice reflections in the neutron diffraction patterns satisfactorily. However, attempt to fit the NPD data with this two phase model led to non-convergence and oscillatory refinement. This was primarily due to the fact that the main Bragg peaks in the pattern is exactly common to both the phases. In view of this, we fitted the NPD pattern using le-Bail approach[14]. As evident from the details of the fit shown in Figure 1a, though the main Bragg peaks are exactely fitted, this model fails to accurately index the peak positions of the weak superlattice peaks. Hence in contrast to the electron diffraction studies, the R3c+P4bm model is not suitable to explain the weak superlattice reflections in this system. We also found that these superlattice reflections could not be exactly indexed on a doubled pseudocubic perovskite cell thereby ruling out the possibility of explaining the structure in terms of the Glazer's classification scheme based on simple tilt systems.[15] This was true for the entire composition range. A need for considering higher order structural modulation therefore arose. By increasing the periodicity equally along the three axes of the cube the best fit was obtained for a 6 x 6 x 6 cell as reported earlier for a KNN free composition.[16] This modulated cell has its volume 216 times the volume of the fundamental cubic cell. We have also attempted to fit the superlattice reflections by choosing modulations of the type $\sqrt{2}$ x $\sqrt{2}$ x n; where n is integer. Such types of modulations have been reported in literature [17-19] and arise due to close proximity of energies associated with polar, antipolar and octahedral tilt distortions .[17] The minimum n required to fit the peak positions accurately in our case was found to be 16 (fig. 1b). This modulated cell has its volume 32 times of the fundamental cell and is considerably smaller than the 6 x 6 x 6 cell reported before.[16]

Fig. 2 shows the polarization and strain response as a function of compositions under different unipolar and bipolar electric fields. For x=0, the shape of the P-E loop is similar to that of a normal ferroelectric (fig 2a). For x=0.01, two anomalies develop: (i) a sudden jump in polarization at 35 kV/cm and (ii) significant constriction of the loop in the middle. Interestingly, during the second cycle and onwards the maximum strain was found to systematically decrease even when the amplitude of field was increased beyond the value in the previous (Fig 3a) cycle. This result is consistant with the bipolar fatigue behavior of 0.01 KNN, reported by Luo et al.[20] The giant strain was however recovered again after heating the specimen at 100 $^\circ$C. This happens because of reversible nature of field induced phase on heating the specimen as reported by Simon et al.[21] for 0.0 KNN. For the next higher composition, x=0.015, the area under the P-E curve is significantly reduced suggesting considerable reduction in the net polarization for similar fields. In contrast to x=0.01, the strain was found to continuously increase with increasing amplitude in the subsequent cycles (Fig. 3b).

Fig. 4 shows the temperature dependence of the relative permittivity of poled and unpoled specimen of x = 0.01 and x = 0.015. As compared to the unpoled specimen, the poled specimen of x=0.01 exhibits a weak but distinctly sharp jump at ~ 50 $^\circ$C on heating. This sharp jump was not found during the cooling cycle (not shown) thereby suggesting that its occurrence in the heating cyle of the poled specimen is manifestation of normal to relaxor transition.

Evidently for x=0.01, room temperature corresponds to non-ergodic relaxor state, and application of high electric field irreversibly increases spatial coherence of the polarization which survives up to ~ 50 $^oC$. XRPD and NPD of the poled specimens of x=0.01 revealed distinct changes as compared to that of the unpoled specimen. While the XRD pattern shows additional peaks flanking the cubic peaks[22], NPD pattern revealed considerable increase in the intensity of superlattice reflections at some 2θ positions and weakening of the intensity at other 2θ positions (Fig. 5). Rietveld analysis proved that the additional peaks in the XRD and NPD correspond to the field induced rhombohedral (R3c) phase[22]. Approximately 34 % R3c phase was found in the poled specimen of x=0.01. It is therefore obvious that the increase in spatial coherence of polarization after poling in the non-ergodic relaxor state is associated with a fraction of the modulated cubic structure transforming to the rhombohedral (R3) phase. It is worth noting that the nature of modulation period remains invariant even after poling.[22] Drastic changes in diffraction patterns have also been reported for NBT[23-27]. In contrast to x=0.01, the next higher composition x = 0.015 did not show signatures of the R3c after poling. The absence of R3c phase after poling x=0.015 can have two interpretations: (i) the cubic to rhombohedral transformation is completely reversible or (ii) the applied field was not sufficient enough to induce the transformation in the first place. The second possibility seems most plausible since the jump in the polarization and strain was not observed for this composition as shown in Fig 2a-2d. Further, the sharp anomaly in the temperature dependence of the dielectric constant of x=0.015 in the heating cycle, corresponding to the normal-relaxor transition, is missing for the poled specimen of x=0.015 (Fig. 4d). It is likely that for this composition the non-ergodic relaxor state occurs below room temperature and hence the room temperature state correspond to ergodic state. It is therefore rather difficult for the electric field to bring about a ferroelectric ordering with rhombohedral structure in this state. We also studied the frequency dependence of the strain and polarization response of the compositions exhibiting giant strain response[22]. A steep increase in the maximum achievable polarization and strain at 40 kV/cm was found to occur below 0.5 Hz (Fig. 6). Since, as already shown above, the sudden increase is associated with field induced cubic to rhombohedral transformation, this study suggests that the minimum critical time desired for the R3c phase to grow appreciably, leading to a giant response is ~ 2 sec. For frequency above 0.5 Hz, the system would not give giant strain response. This observation contradicts the results of other groups[28,29]where they have observed field induced phase transformation can occur for a wide frequency range for some other NBT based systems.

In conclusion we have shown that the giant piezo strain in the lead-free system (0.94-x)$Na_{0.5}Bi_{05}TiO_3$-$0.06BaTiO_3$-$xK_{0.5}Na_{0.5}NbO_3$ is associated with the system's transformation from a structural state with $\sqrt{2}$ x $\sqrt{2}$ x 16 type structural modulation to a rhombohedral (R3c) phase at a critical field in the non-ergodic relaxor state. Partial irreversibility of field induced R3c phase in the first cycle led to considerable reduction in strain in second and subsequent cycles. The giant strain could be recovered again only after annihilation of the quenched R3c phase by tasking it into the ergodic state (~50 $^{o}$C).

Acknowledgment: RR is grateful to Science and Engineering Board of the Department of Science and Technology for financial assistance (Grant number : SERB/F/5046/2013-14)

Figures:

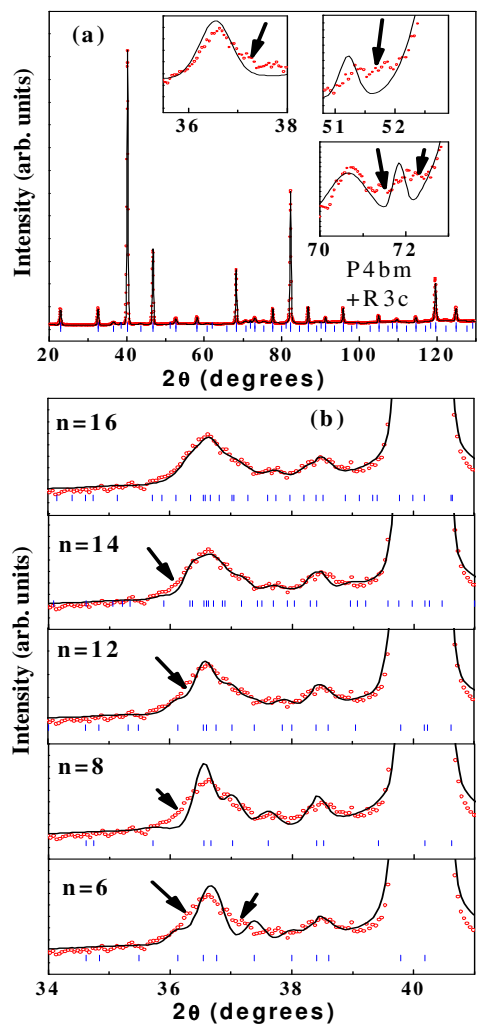

FIG.1. (a) (Color Online) leBail fitting of neutron powder diffraction pattern of unpoled 0.93NBT- 0.06BT-0.01KNN with R3c+P4bm model. Inset shows the 2θ regions exhibiting misfit between the observed and the calculated superlattice reflections (b) leBail fitting of the NPD pattern with √2 x √2 x n type supercells.

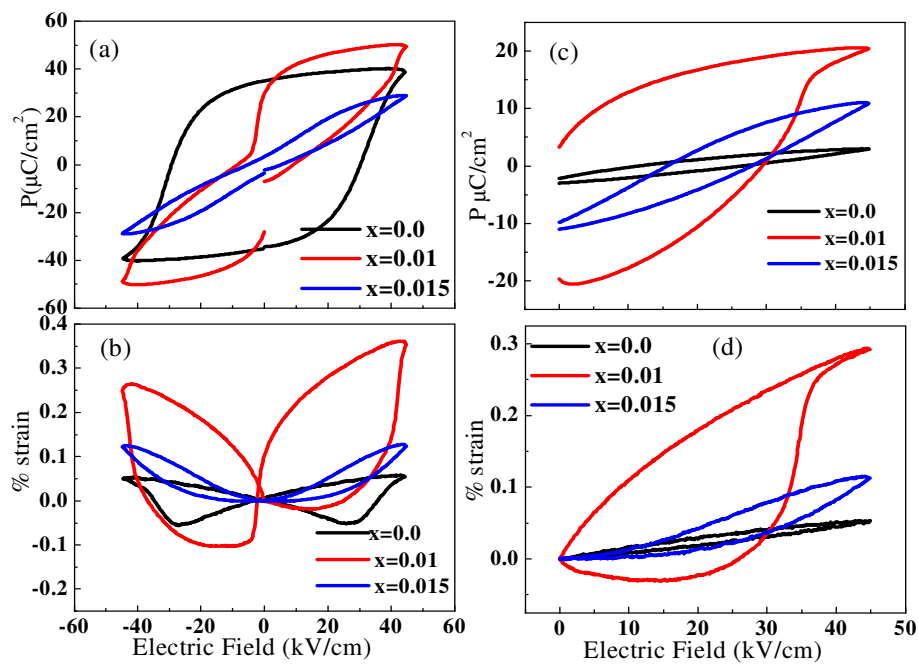

FIG.2. Electric field dependent polarization and strain measurement for different NBT-BT-KNN compositions at an electric field of 45 kV/cm. (a) bipolar P-E, (b) bipolar S-E, (c) unipolar P-E and (d) unipolar S-E.

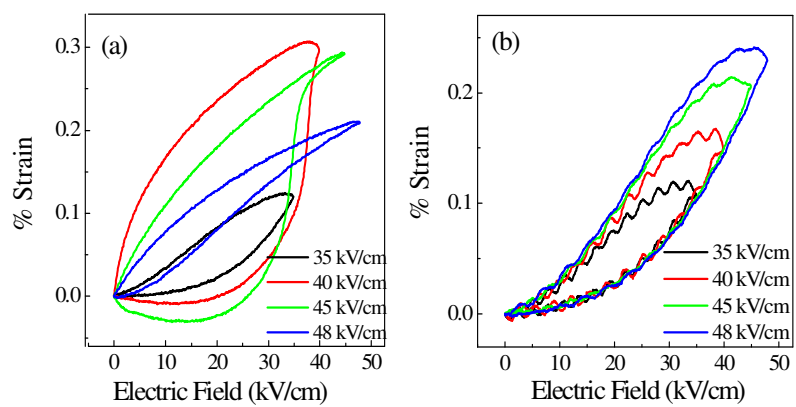

FIG.3. Electric field dependent unipolar strain of (a) 0.01 KNN and (b) 0.025KNN.The amplitude of the field was increased in subsequent cycles for the case (b).

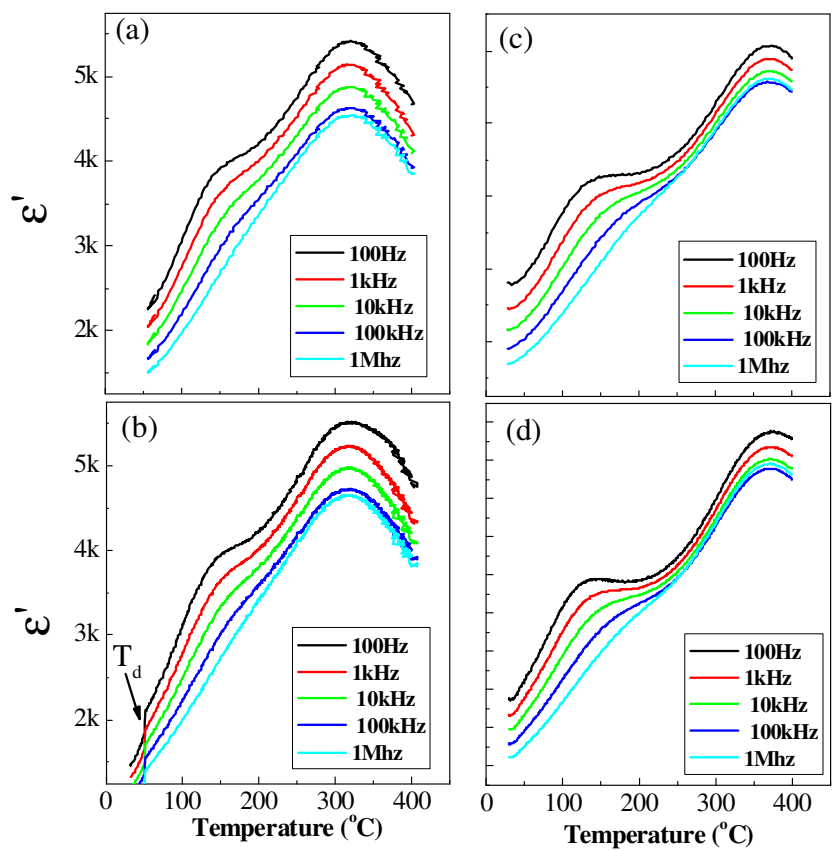

FIG.4. Temperature-dependent permittivity of NBT-BT-KNN with (a) 0.01KNN-unpoled, (b) 0.01KNN-poled, (c) 0.015KNN-unpoled and (d) 0.015KNN-poled.

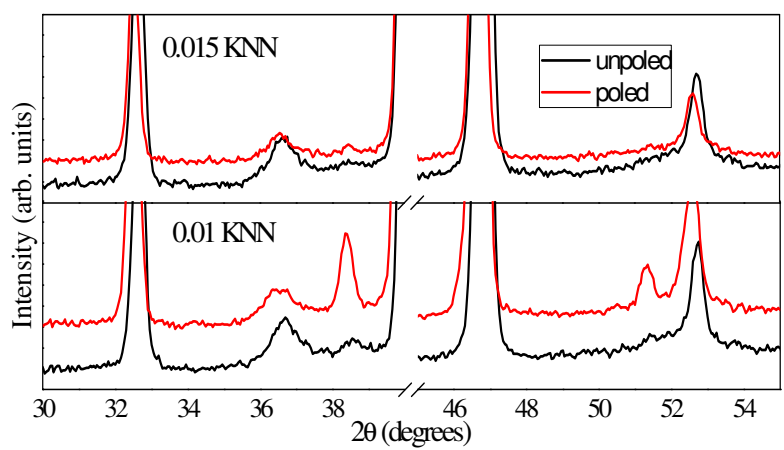

FIG.5. (Color Online) Neutron powder diffraction patterns of poled and unpoled NBT-BT-KNN for x=0.01 (bottom panel) and x=0.015 (top panel).

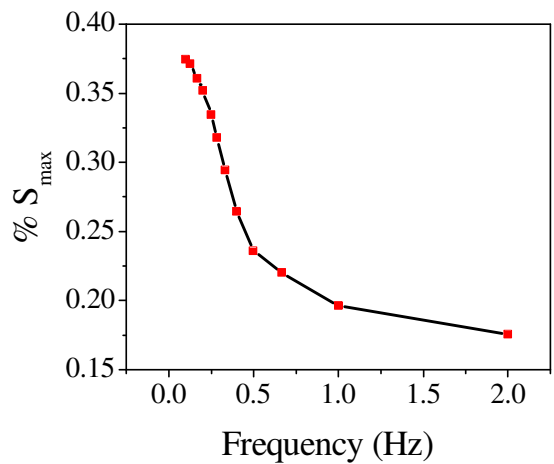

FIG.6. Frequency dependence of maximum strain measured at 40 kV/cm for 0.005 KNN.